\documentclass{aanew}
\usepackage{epsfig,graphicx}
\begin{document}


\title{On the nature of the hard X-ray source \object{4U\,2206+54}}

\author{Ignacio~Negueruela\inst{1,2,3,4}               
\and Pablo~Reig\inst{5,6}}

\institute{Observatoire de Strasbourg, 11 rue de l'Universit\'{e},
F67000 Strasbourg, France
\and SAX SDC, Agenzia Spaziale Italiana, c/o Telespazio, via Corcolle
19, I00131 Rome, Italy
\and Astrophysics Research Group, Liverpool John Moores University,
Byrom St., Liverpool, L3 3AF, U.K.
\and Physics and Astronomy Department, Southampton University,
Southampton, SO17 1BJ, U.K.
\and Foundation for Research and Technology-Hellas, GR711 10, Heraklion,
Crete, Greece
\and Physics Department, University of Crete, GR710 03 Heraklion, Crete, Greece}

\offprints{I. Negueruela, Strasbourg address, \email{ignacio@astro.u-strasbg.fr}}

\date{Received /Accepted     }

\abstract{
The recent discovery of a $\sim 9.5\:{\rm d}$ period in the X-ray lightcurve
of the massive X-ray binary \object{4U\,2206+54} has opened the
possibility that it is
a Be/X-ray binary with an unusually close orbit, which, together with
its low intrinsic luminosity, suggests that the system is actually
a Be + WD binary, in which a white dwarf accretes material from
the dense circumstellar disc surrounding a classical Be star. In this 
paper we present new X-ray observations and for the first time 
high-resolution optical spectroscopy of the source. We show that
both the X-ray behaviour and the characteristics of the optical
counterpart, \object{BD$\:+53^{\circ}\,$2790}, are more consistent
with a neutron star accreting from the wind of an early-type star. The
X-ray lightcurve shows irregular flaring and no indications of
pulsations, while the very high hydrogen column density supports
accretion from a dense wind. \object{BD$\:+53^{\circ}\,$2790} is
shown not to be a classical Be star, as believed until now, but rather
a very peculiar late O-type active star, exhibiting emission
components in the \ion{He}{ii} lines, complex spectral variability
and strong wind resonance lines in the ultraviolet. Though many of
the characteristics of the spectrum resemble those of the He-rich
stars, the absence of \ion{He}{i} variability makes a connection
unlikely. The spectrum is compatible with a composite of two stars of
similar spectral type, though circumstantial evidence points to a
single very peculiar active early-type star. This adds weight to the
growing evidence that the traditional subdivisions of  supergiant and
Be/X-ray binaries fail to cover the whole phenomenology of massive
X-ray binaries. 
}

\maketitle        
\titlerunning{The nature of \object{4U\,2206+54}}               
\authorrunning{Negueruela \& Reig}

\section{Introduction}

High Mass X-ray Binaries (HMXBs) are X-ray sources composed of an early-type 
massive star and an accreting compact object (generally a neutron star, but 
occasionally a black hole and, at least theoretically, possibly a white
dwarf). HMXBs are traditionally divided (see Corbet 1986) into Classical or
Supergiant X-ray  binaries (SXBs) in which the compact object 
accretes from the 
stellar wind (sometimes directly from the atmosphere through localized
Roche-lobe overflow) of an OB supergiant and Be/X-ray binaries (BeXBs), 
in which a neutron star orbits an unevolved OB star surrounded by a dense 
equatorial disc. Almost all known HMXBs fit well into one of these two 
categories (with a majority of systems being BeXBs), though a few 
systems, such as \object{LMC X-4} \cite{hutal78} or \object{RX
  J1826.2$-$1450} \cite{mot97}, seem to contain a compact
object accreting from a ``normal'' main-sequence O-type 
star.

The hard X-ray source \object{4U\,2206+54} was first detected by 
the {\em Uhuru} satellite \cite{gia72}. It appeared
in the {\em Ariel V} catalogue as \object{3A\,2206+543} \cite{waral81}.
Steiner et al. (1984; hereafter S84) used the refined position from 
the {\em HEAO--1} Scanning Modulation Collimator to identify the optical
counterpart with the early-type star
\object{BD$\:+53^{\circ}\,$2790}. S84 reported that the 
H$\alpha$ line was in emission, showing two distinctly separated peaks
with $\Delta v_{{\rm peak}}= 460\:{\rm km}\,{\rm s}^{-1}$. From their
photometry, they estimated that the counterpart was a B0--2e main sequence 
star, and therefore concluded that the system was a Be/X-ray binary. 
In this subclass of HMXBs, 
the X-ray emission is due to accretion of matter from a Be star
by a compact companion (see Bildsten et al. 1997; Negueruela 1998). The 
name ``Be star'' is used as a general term describing an early-type luminosity
class III--V star, which at some time has shown emission in the Balmer series 
lines (Slettebak 1988, for a review). Both the emission
lines and the characteristic strong infrared excess when compared to
normal stars of the same spectral types are attributed to the presence
of circumstellar material in the shape of a decretion quasi-Keplerian
disc (see Negueruela \& Okazaki 2000 for a recent discussion).

Assuming a distance to \object{4U\,2206+54} of 2.5 kpc, S84 calculate
an average luminosity for the source of $L_{{\rm x}} \simeq
7\times10^{34}$ erg  s$^{-1}$ between 1974 November and 1981 October.
Saraswat \& Apparao (1992, henceforth SA92) presented X-ray
observations of \object{4U\,2206+54} made with the {\em EXOSAT}
satellite at different epochs between 1983--1985. The source was always
detected, though in different states. In August 1983 and June 1985, the 
source was active, with a low-level 
luminosity of $\approx 5\times 10^{34}$ erg s$^{-1}$ and 
aperiodic flaring phases (a few hundred seconds long) in which the
overall X-ray flux increased by a factor 3\,--\,5 and the X-ray spectrum  
changed, becoming harder. In December 1984, the source was in
quiescence, and the X-ray flux was weak 
($L_{{\rm x}} \approx 3\times 10^{33}$erg s$^{-1}$) 
and stable. SA92 also announced the possible detection of a spin period for 
the compact object which would be in the range 390\,--\,400 s and suggested
that the accreting object was a white dwarf.

The source appears in the {\em ROSAT} All Sky Survey \cite{vog99} as 
\object{1RX$\:$J220755+543111} and has been consistently detected by
the All Sky Monitor on board {\em RXTE} according to the quick-look
results provided by the ASM/{\em RXTE} team. 
Corbet et al. (2000) have announced the detection
of a $9.570\pm0.004\:{\rm d}$ periodicity in the X-ray lightcurve. If
this is the binary period, then it would be the shortest known for a 
BeXB -- unless the $\sim1.4\:{\rm d}$ periodicity in the optical
lightcurve of \object{RX$\:$J0050.7$-$7316} reflects its orbital period 
\cite{CO00}.

\section{Observations}

 \object{BD$\:+53^{\circ}\,$2790} (= \object{LS\,III +54$^{\circ}$16}
 = \object{Hilt 1086}) is  
included in several catalogues of bright stars. Measurements of its optical 
magnitudes are reported since the work of Hiltner \& Johnson (1956). In spite
of this, very little previous work on this source has been reported. We have
undertaken a major multi-wavelength monitoring campaign on this source, the
results of which will be presented in a subsequent paper. Here
we concentrate only on observations that provide information on the
 nature of the system.

\begin{figure}
\resizebox{\hsize}{!}{\includegraphics{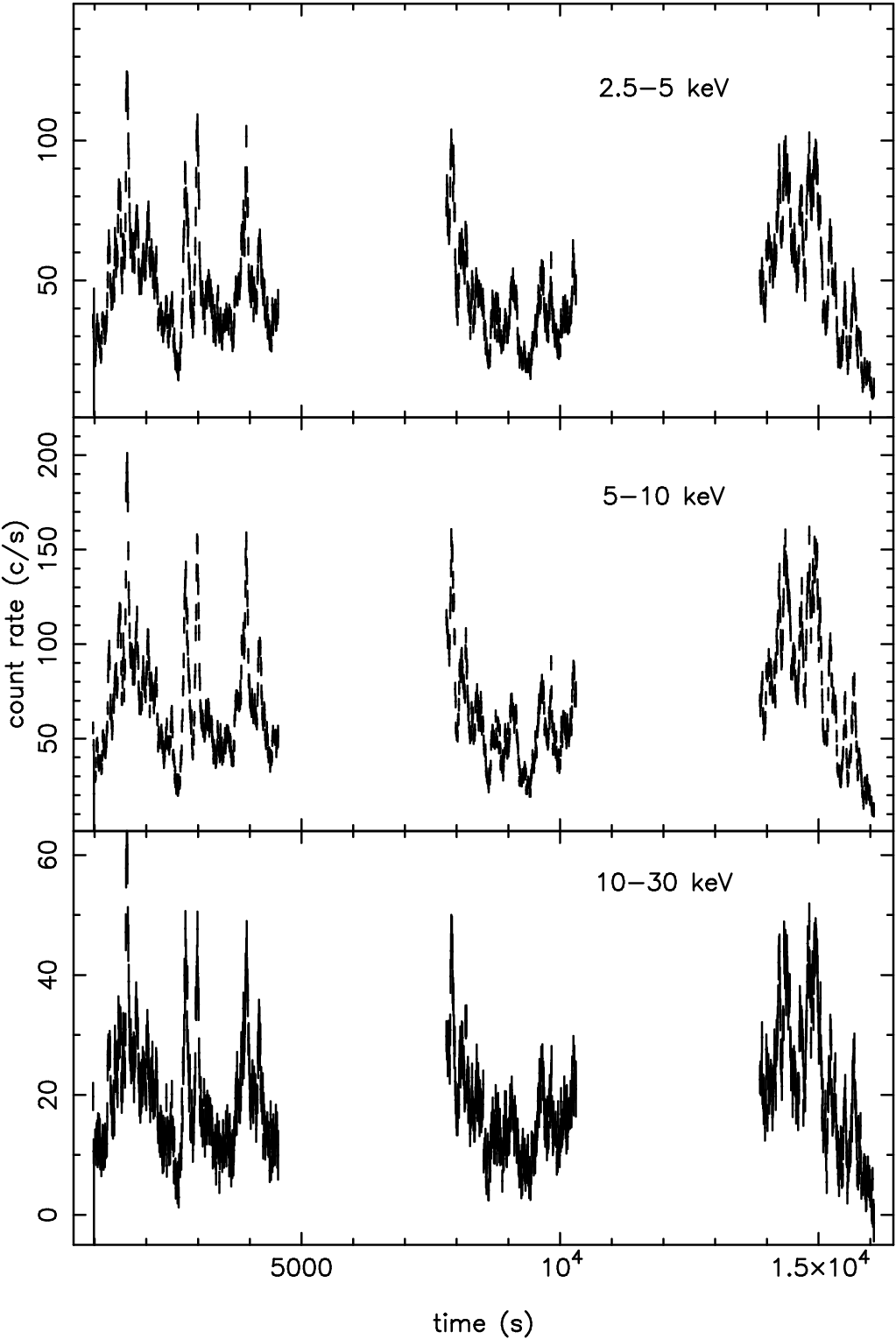}}
    \caption{Background subtracted {\em RossiXTE} lightcurves in three energy
ranges. The original lightcurves have been rebinned into 8-s bins. The
light-curve, clearly dominated by flaring activity, is similar to 
those of low X-ray luminosity supergiant binaries.}
    \label{fig:xraylc}
\end{figure}

\subsection{X-ray Observations}

We have analysed X-ray data taken with the Proportional Counter Array
(PCA) onboard  the {\em Rossi X-ray Timing Explorer} ({\em RXTE}). 
The data were
retrieved from the {\em RXTE} archive and correspond to an observation made on
March 11\,--\,13, 1997. After the screening and filtering of data, i.e., 
ensuring
that all five PCA units were  functioning and removing data taken at low
Earth elevation angle ($<$  10$^{\circ}$) and during times of high particle
background, we were left with $\sim$ 9000 s of on-source clean data.
Also in order to improve signal-to-noise we selected only events from the
top layer (the PCUs have three Xenon layers, each consisting of two anode
chains; see Jahoda et al. 1996 for a technical description of the instrument).

Figure~\ref{fig:xraylc} shows the background subtracted light curves of
 \object{4U\,2206+54} in three
different energy ranges.  The temporal variability is characterised by
erratic flaring activity on short timescales. The intensity shows changes
by a factor of 3 in less than 2 minutes. The source becomes increasingly 
variable as the energy increases. The
$rms$ of the light curves varies from $\sim$ 40\% for the energy range
2.5\,--\,5 keV to $\sim$ 45\% for 5\,--\,10 keV and $\sim$ 50\% for 
10\,--\,30 keV. This flaring 
and erratic behaviour was also reported by SA92 during their observations
(which covered the 2\,--\,10 keV range) on the two occasions in which
the source was active. Similar lightcurves are observed in SXB
binaries in which a neutron star accretes from the radiative wind of
an evolved star, such as 
 \object{2S\,0114+65} \cite{yamal90} or \object{Vela X-1}
 \cite{kre99}, which have typical $L_{{\rm x}} \sim 10^{35} -
 10^{36}\:{\rm erg}\,{\rm s}^{-1}$ (from now on, low-luminosity SXBs).  

\begin{figure}
\resizebox{\hsize}{!}{\includegraphics{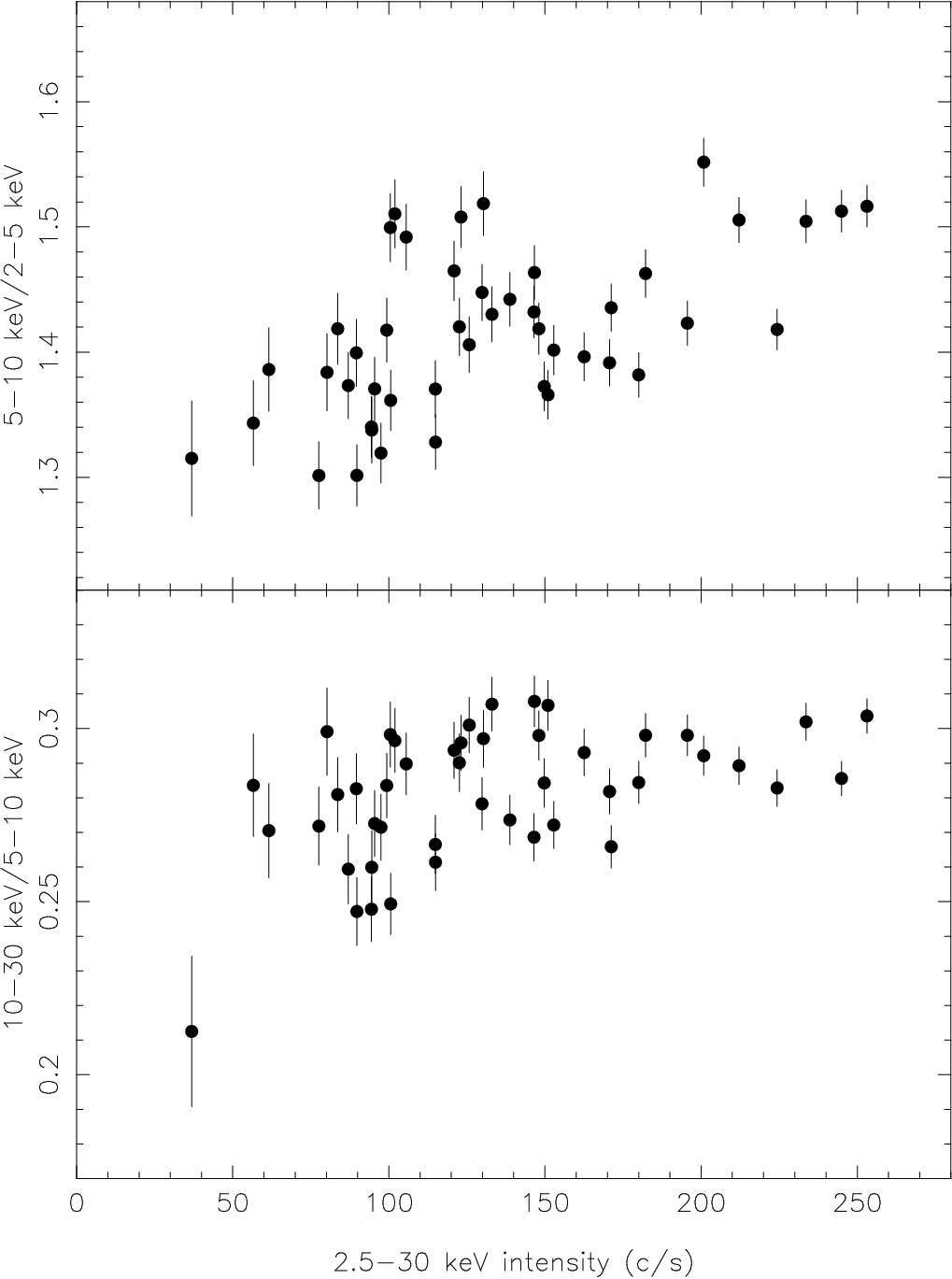}}
    \caption{
Hardness ratios as a function of 2.5\,--\,30 keV count rate. The 
X-ray spectrum
tends to become harder as the count rate increases.}
    \label{fig:xrayhardness}
\end{figure}

\begin{table}
\begin{center}
\caption{Spectral fit parameters and 68\% confidence errors. The fit was
performed in the energy range 2.5--30 keV}
\begin{tabular}{lc}
\hline
$N_{\rm H}$ (10$^{22}$ atoms cm$^{-2}$)       &  $4.7\pm0.2$  \\
$\Gamma$                                &  $1.71\pm0.3$ \\
$E_{\rm cut}$ (keV)                         &  $7.4\pm0.2$ \\
$E_{\rm fold}$ (keV)                        &  $17.5\pm0.8$  \\
normalization                           &  $0.085\pm0.003$ \\
$\chi^2_{\rm r}$(dof)                         &  0.9(56)     \\
\hline
\end{tabular}
\end{center}
\label{tab:fitspec}
\end{table}

The correlation between the hardness ratio 5\,--\,10 keV/2.5\,--\,10 keV 
and the
count rate (see Fig.~\ref{fig:xrayhardness}) indicates that the X-ray
 spectrum becomes harder
during the peak of the flares. 

\begin{figure}
\resizebox{\hsize}{!}{\includegraphics{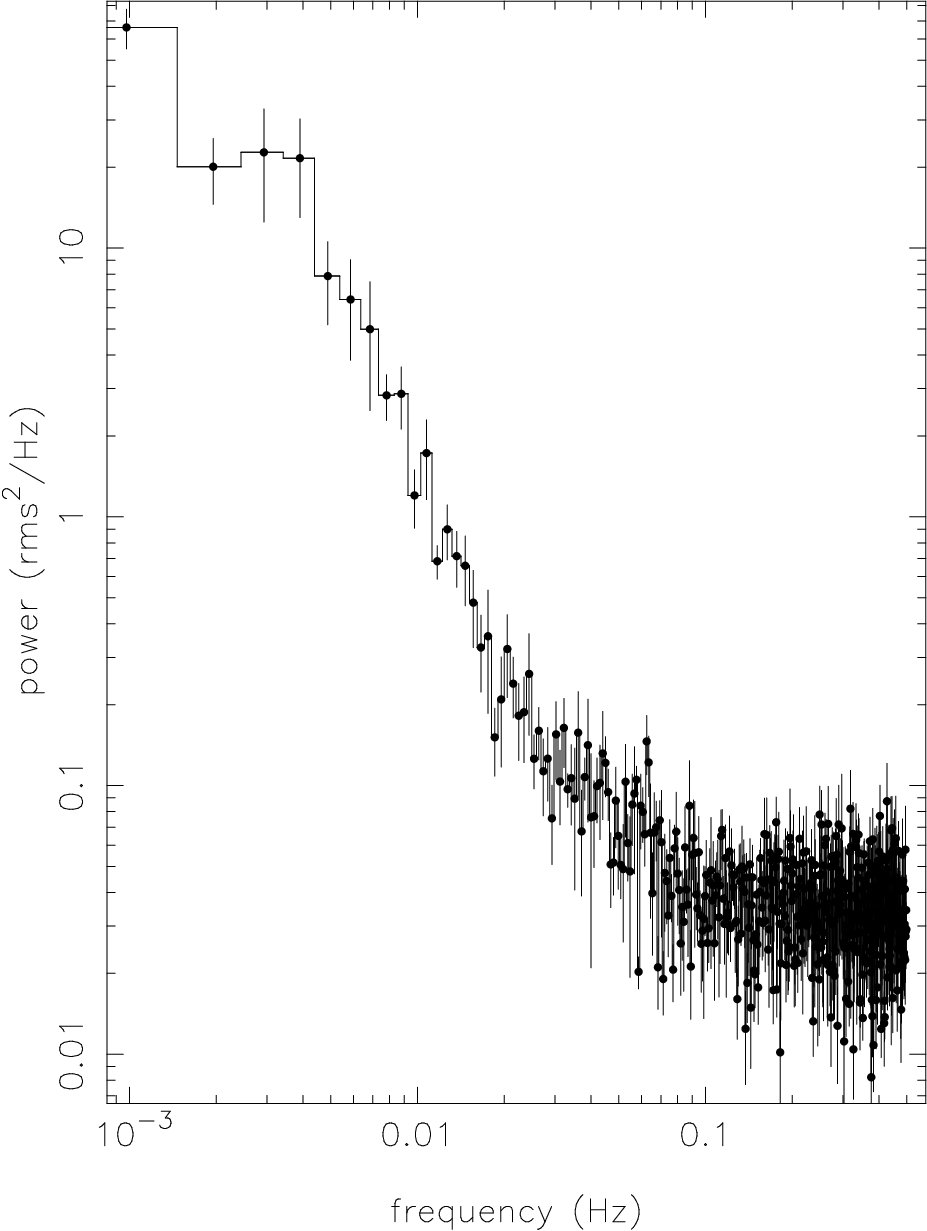}}
    \caption{Power spectrum for the {\em RXTE}/PCA observation of
      \object{4U\,2206+54}.} 
  \label{fig:period}
\end{figure}

Figure~\ref{fig:period} shows the power spectrum of
\object{4U\,2206+54}. No evidence 
for the 390\,--\,400 s pulse period reported by SA92 was found. The power 
spectrum is dominated by a strong red noise component and no periodicity
is detected at any significative level.

\begin{figure}
\resizebox{\hsize}{!}{\includegraphics{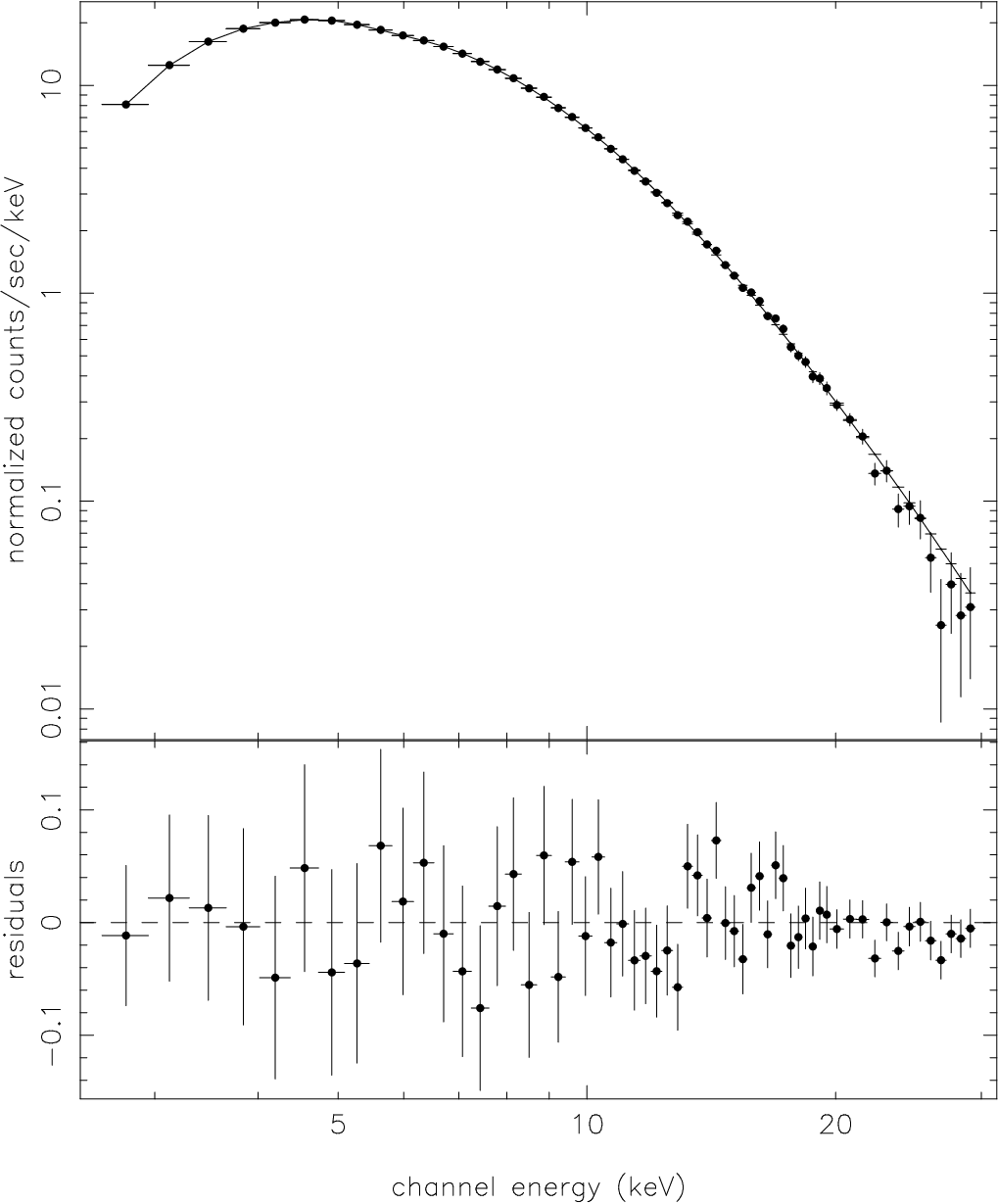}}
    \caption{X-ray spectrum of \object{4U\,2206+54} (circles) and the best-fit
model (solid line). The continuum is represented by a power-law plus a 
cutoff at 7.4 keV (see Table~1). No iron line is required.}
  \label{fig:xrayspectrum}
\end{figure}

Spectral analysis was performed on {\em Standard2} data, in the energy
range 2.5\,--\,30.0 keV. The best-fit model was an absorbed power-law and 
a high
energy cutoff yielding an unabsorbed X-ray flux of $4.8 \times 10^{-10}\:
{\rm erg}\,{\rm cm}^{-2}\,{\rm s}^{-1}$. The best-fit parameters and their 
68\% confidence
errors are given in Table~1, while
Fig.~\ref{fig:xrayspectrum} shows the photon distribution. No evidence
for an iron line at around 6.4 keV was found. This line is seen in the
spectra of the low-luminosity SXBs which display similar X-ray
lightcurves. We can set an upper limit on the equivalent width of such
line at $< 10$~eV.

\subsection{Ultraviolet spectroscopy}

\begin{figure*}
  \begin{picture}(500,360)
\put(0,0){\includegraphics{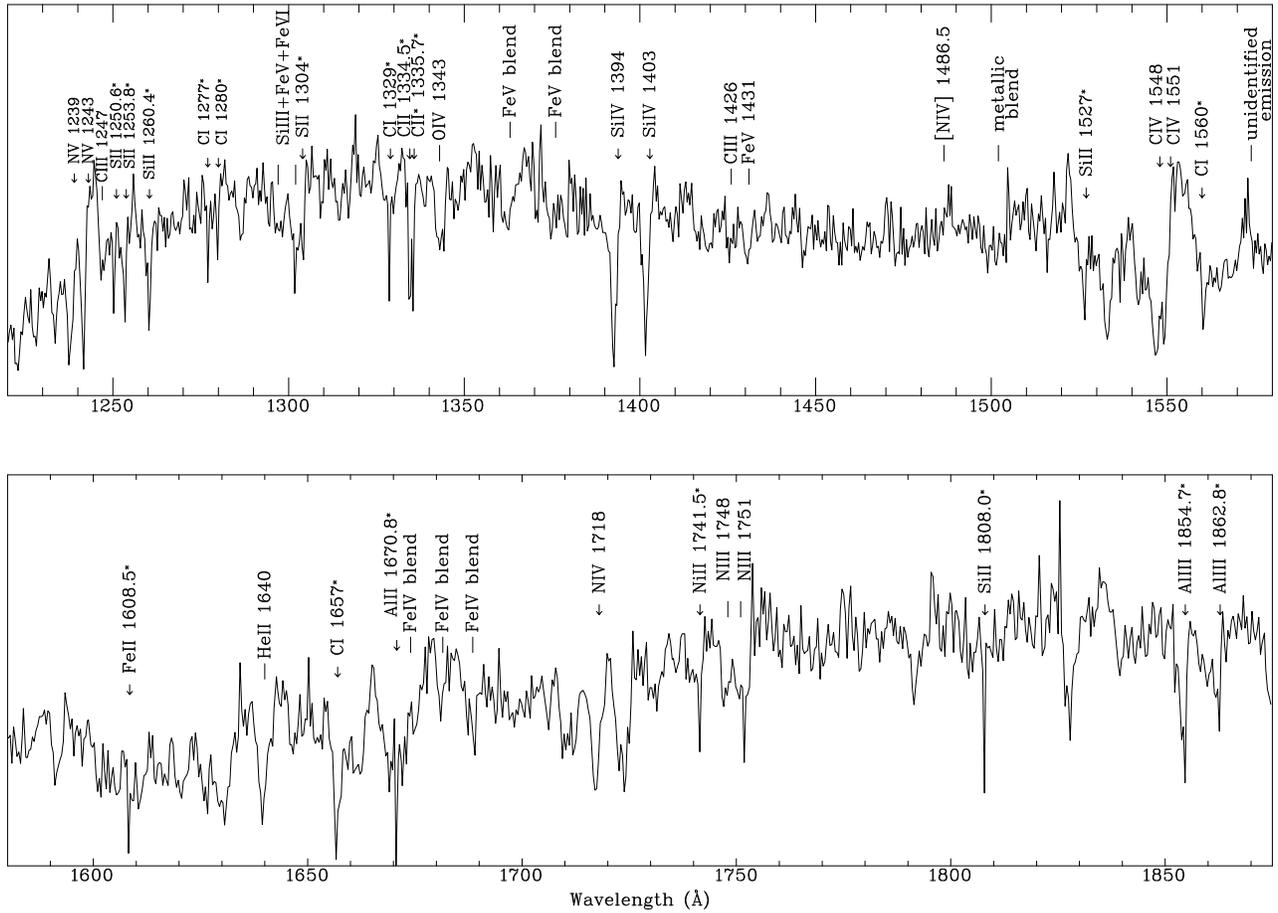}}
\end{picture}
    \caption{The ultraviolet spectrum of
      \object{BD$\:+53^{\circ}\,$2790}. The arrows
indicate the rest wavelength of the interstellar lines (marked with
`$\,^{*}$') and the wind lines N\,{\sc v} $\lambda \lambda$1239,
1243 \AA, Si\,{\sc iv} $\lambda \lambda$1393.8, 1402.8 \AA, C\,{\sc iv}
$\lambda \lambda$1548.2, 1550.8 \AA\ and N\,{\sc iv} $\lambda$1718\AA.}
    \label{fig:uvspectrum}
\end{figure*}

Ultraviolet observations of \object{BD$\:+53^{\circ}\,$2790}  were
retrieved from the {\em International Ultraviolet Explorer} archive at
Rutherford Appleton Laboratory. The
database contains pre-processed spectra, which were subsequently reduced and
analysed using the {\em Starlink} packages {\sc iuedr} \cite{gidal96}
and {\sc dipso} \cite{how}. The low resolution
spectra from the Uniform Low Dispersion Archive (ULDA) LWP18128 and
SWP39111 did not provide enough detail to 
allow accurate line identification. The high resolution spectrum
SWP39112, taken with the short-wavelength camera in the large
aperture mode on June 18, 1990, is displayed in
Fig. \ref{fig:uvspectrum}. The original resolution of the spectrum is
approximately 0.05 \AA, but it has been rebinned to 0.4
\AA\ for display. The wavelength calibration, which has been checked
with different interstellar lines, is accurate to a few km s$^{-1}$.

The most remarkable features are the strong P-Cygni profiles of the
resonance wind doublets C\,{\sc iv} $\lambda \lambda$1548.2, 1550.8
\AA\ and N\,{\sc v} $\lambda \lambda$1239, 1243 \AA. They are
stronger than those reported for Be stars of different spectral types
by Prinja (1989) and resemble those typical of O-type main
sequence stars \cite{walal85}. The
subordinate wind line N\,{\sc iv} $\lambda$1718 \AA, which generally
follows the behaviour of the resonance doublets \cite{WP84a}, does not
show clear evidence for a P-Cygni profile, but it
could be masked by the blend of two strong metallic lines just
shortwards of it. The Si\,{\sc iv} $\lambda \lambda$1394, 1403~\AA\
doublet shows only a moderate wind effect, with narrow absorption troughs.

\subsection{Optical Spectroscopy}

The source has been monitored since 1990, using a large array of telescopes 
and configurations.
The complete dataset, its variability and searches for periodic
behaviour will be reported in a forthcoming paper.
Here we concentrate on relatively high-resolution blue spectroscopy taken
with the 2.5-m Isaac Newton Telescope (INT), located at the Observatorio 
del Roque de los Muchachos, La Palma, Spain, on July 11, 1995 and 
August 3, 1998. Both observations
were taken with the Intermediate Dispersion Spectrograph (IDS) equipped
with the 235-mm camera and the R1200B grating. In 1995, the Tek3 CCD
was in use, while in 1998, the EEV\#12 CCD had replaced it. Further
blue observations were taken on July 25\,--\,30, 2000, using the 1.52-m
G.~D.~Cassini telescope at the Loiano Observatory,
Italy, equipped with the Bologne Faint Object Spectrograph and Camera
(BFOSC). Several observations were taken using grism\#6, while two
higher resolution spectra were taken with grism\#9 in echelle mode
(using grism\#10 as cross-disperser). 

All the data have been reduced using the {\em Starlink}
software packages {\sc ccdpack} \cite{dra98} and
{\sc figaro} \cite{sho97}  and
analysed using {\sc dipso}.

\section{The optical/UV spectrum}

\begin{figure*}
\begin{picture}(500,220)
\put(0,0){\includegraphics{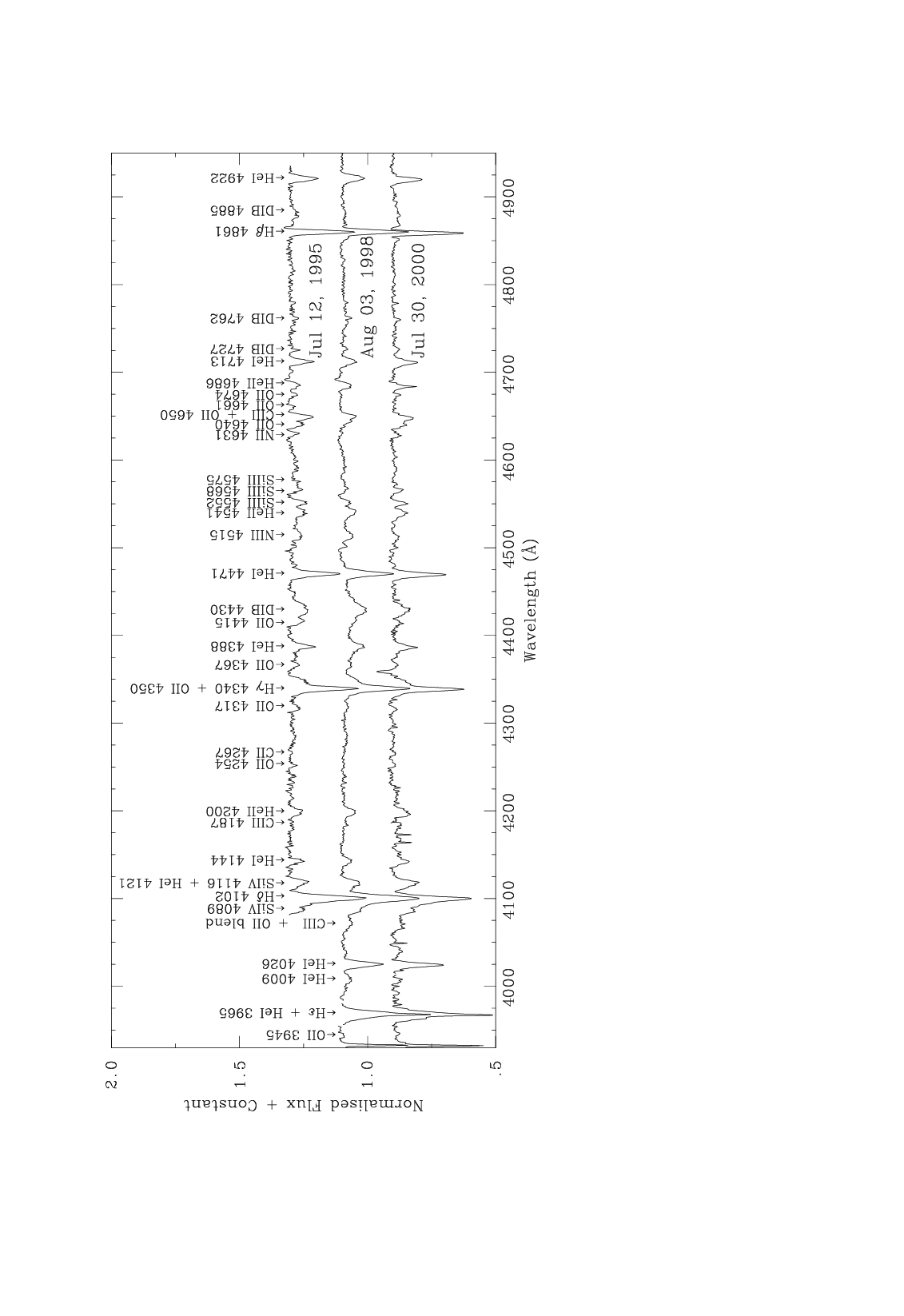}}
\end{picture}
 \caption{High-resolution blue spectra of \object{BD$\:+53^{\circ}\,$2790} 
showing spectral variability. All spectra have been divided by a 
spline fit to the continuum for 
normalization and offset by a constant amount. Diffuse Interstellar
Bands (DIBs) are indicated.}
    \label{fig:bluespectrum}
\end{figure*}

The spectra of \object{BD$\:+53^{\circ}\,$2790} in the classification
region do not readily correspond to any spectral type. This fact was
recognised by Hiltner \& Johnson (1956), who classified the star as
O9IIIp from photographic plates (apparently not detecting any emission
in the blue at the time). An immediate conclusion of our monitoring
(see Fig.~\ref{fig:bluespectrum}) is that the spectrum is also {\em
variable}.  The presence of 
strong He\,{\sc ii} lines, specially He\,{\sc ii} $\lambda$4200 \AA, would
indicate an O-type classification. Though the spectrum from August 1998
is relatively close to that of a normal O9 star with high $v\sin i$,
most other blue spectra of the source (such as those from July 1995 and
July 2000) display abundant and strong O\,{\sc ii} lines, together with
a relatively strong Si\,{\sc iii} triplet. These lines do not correspond
to an O-type star, but are typical of early B-type stars.

From our spectra, we can deduce the following information:

\begin{itemize}
\item \ion{He}{ii} $\lambda$4686\AA\ is clearly variable, has obvious
emission infilling and probably a permanent P-Cygni-like emission
component.
\item \ion{He}{ii} $\lambda$4200\AA\ is possibly also variable, though not
to the same extent as \ion{He}{ii} $\lambda$4686\AA.
\item All \ion{O}{ii} and \ion{Si}{iii} lines are variable. They can
be basically absent, as in the August 1998 spectrum, or rather strong,
as in July 1995, with the July 2000 spectrum showing an intermediate state.
Other metallic lines, such as \ion{N}{ii} $\lambda$4631\AA\ and
\ion{C}{ii} $\lambda$4267\AA\ seem to participate in these changes, though 
\ion{N}{iii} $\lambda$4515\AA\ apparently does not. All this would indicate
a lower-temperature component superimposed on the spectrum of an O9 star.
\item H$\beta$, like H$\alpha$ and the \ion{He}{i} lines at 
$\lambda\lambda$ 6678, 7065 \AA\ displays a shell-like spectrum, in the sense
that the absorption feature is narrow and there are variable emission
components in the wings. \ion{He}{i} lines in the blue do not show
this phenomenology.
\item The upper Paschen series is basically in absorption, unlike in all
the Be/X-ray binaries observed (see Negueruela \& Torrej\'{o}n, in 
preparation).
\end{itemize}

The ultraviolet spectrum is typical of a main-sequence, late O-type star.
The O-star classification is supported by the strength of the Fe\,{\sc v} 
lines in the ultraviolet (specially the ratio Fe\,{\sc v} $\lambda$1431$/$
C\,{\sc iii} $\lambda$1427 $>$ 1; Walborn et al. 1985) and the strong 
\ion{He}{ii} $\lambda$4200\AA\
and \ion{N}{iii} $\lambda$4515\AA\
in the blue \cite{WF90}, none of which would 
be expected at the temperature of an early
B-type star. The luminosity classification is supported by the shallow
and broad photospheric lines in the ultraviolet and fundamentally the 
lack of wind troughs in the Si {\sc iv} $\lambda \lambda$1394, 1403~\AA\
resonance lines, which show a very strong
luminosity dependence \cite{WP84b}.

The photospheric lines are those characteristic of a late-O/early-B
main-sequence star, with Fe\,{\sc v} lines dominating the spectrum
shortwards of $\lambda$1500 \AA. The Al\,{\sc iii} $\lambda
\lambda$1854.7, 1862.8 \AA\ lines, although dominated by the sharp
interstellar features seem to have shallow broader 
photospheric components, which
would indicate a spectral type not much earlier than B0. The
photospheric lines seem, in general, to be broader than those of the
standards listed by Walborn et al. (1985).

\begin{figure*}[ht]
\begin{picture}(500,220)
\put(0,0){\includegraphics{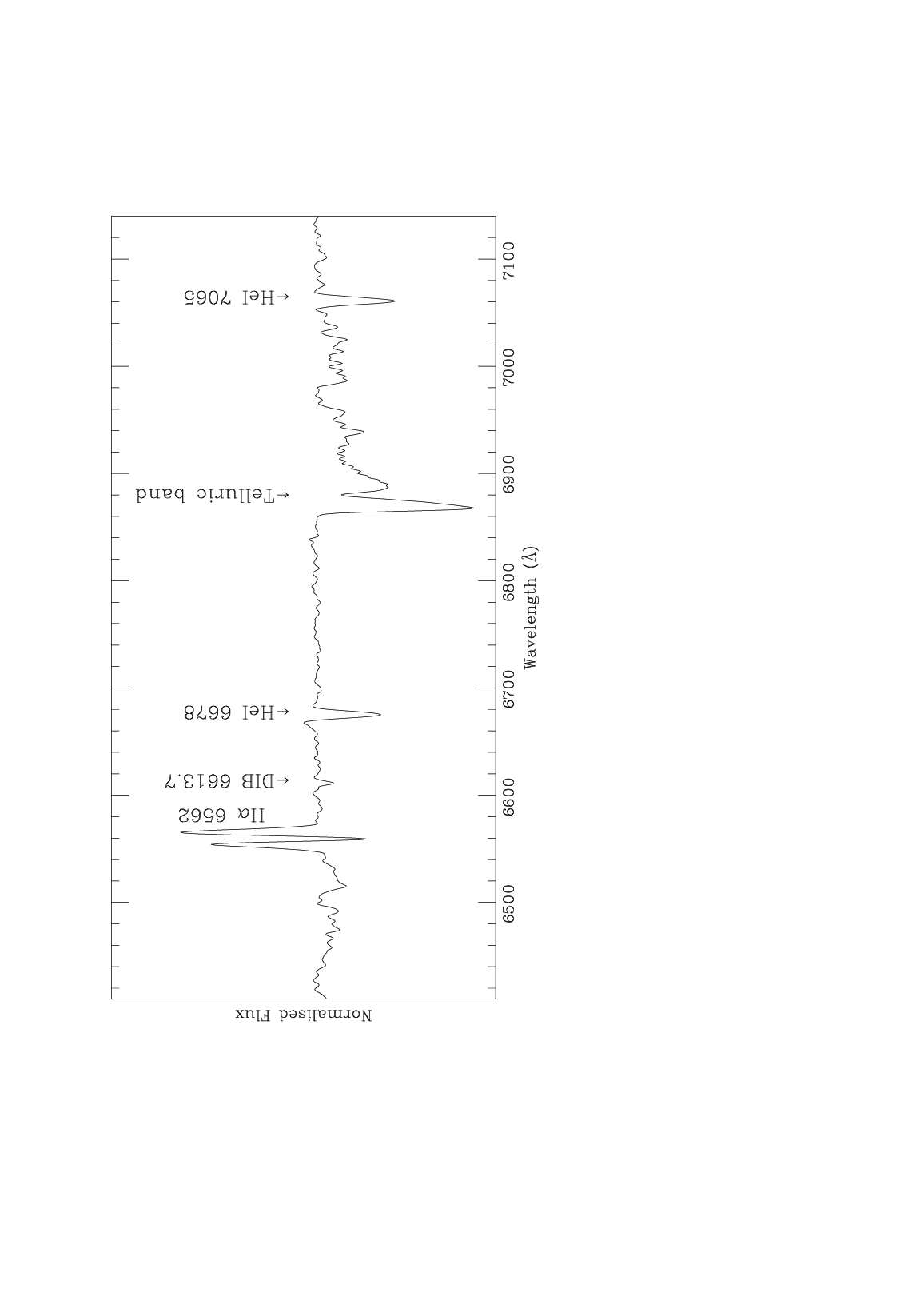}}
\end{picture}
\caption{Red spectrum of \object{BD\,+$53^{\circ}$2790} (one order of
 the echelle spectrum taken on Jul 29, 2000, from
the 1.52-m G.~D.~Cassini Telescope) showing the typical characteristics
of the spectra of the source in that region: 
shell-like features in H$\alpha$ and He\,{\sc i} lines, R$>$V asymmetry
in H$\alpha$ and the inverse symmetry in He\,{\sc i} $\lambda6678$\AA.
The spectrum has been divided by a 
polynomial fit to the continuum for 
normalisation and smoothed with a $\sigma=0.3$\AA\ Gaussian.} 
\label{fig:redspectrum} 
\end{figure*}

In a normal star, the condition 
\ion{He}{ii}~$\lambda$4545\AA~$\simeq$~Si\,{\sc iii} $\lambda$4552\AA\
would indicate a spectral type O9.5. The ratio between
\ion{He}{i}~$\lambda$4471\AA\ and \ion{He}{ii}~$\lambda$4545\AA, which
remains constant (He lines do not change intensity), also 
indicates a spectral type O9.5 (using the values from Mathys
1988). On the other hand, the presence of a variable spectral component, 
corresponding to a lower effective temperature, cannot be readily
reconciled with the idea of a normal single O-type star. The following
possibilities are open:

\subsection{BD\,$+53^{\circ}$2790 as a peculiar Oe shell star}

The shapes of the H$\alpha$ and He\,{\sc i} lines in the red are, at first 
sight, typical of a shell star. Shell lines are believed to be formed when 
the line of sight to the observer is intercepted by the outer cooler parts of 
the envelope of a Be star, which absorb the photospheric continuum
(see Hanuschik 1995, 1996). Therefore Be stars seen very close to edge-on
show deep absorption cores going down below the continuum level on top 
of their emission lines and are referred to as shell stars.

The spectra of Be shell stars are characterised by an absorption 
spectrum corresponding to a lower temperature (generally displaying 
many weak lines corresponding to \ion{Fe}{ii} and other singly-ionised 
metals) superimposed on the photospheric spectrum. To the best of our 
knowledge, the spectrum of an Oe shell star
has never been described in the literature. It is therefore possible to
speculate that the envelope surrounding an Oe shell star could produce
the \ion{O}{ii} and \ion{Si}{iii} lines, which correspond to a 
temperature of $\sim 25000\:{\rm K}$.

This would not explain the emission component almost certainly seen in 
\ion{He}{ii} $\lambda$4686\AA\ and would imply that either 
\object{BD\,$+53^{\circ}$2790} is both an Of and an Oe star (the first ever
identified) or that the \ion{He}{ii} emission originates in an 
accretion disc around the compact object. The presence of an accretion
disc in the system, though not directly ruled out by observations, is
difficult to reconcile with the absence of photometric variability.
For example,  Hiltner \& Johnson (1956) give $U=9.40$, $B=10.11$, 
$V=9.86$. Barbier et al. (1973) report $U=9.46$, $B=10.12$, $V=9.82$, 
while S84 give 
$U=9.45$, $B=10.05$, $V=9.85$, $R=9.61$, $I=9.41$ measured with the 0.91-m 
telescope at Kitt Peak National Observatory on October 5, 1981. Given the 
expected errors (for instance, St84 estimate their errors at 0.04 mag), the 
values of $UBV$ are consistent with no variation at all. The Tycho
catalogue gives magnitudes $B_{{\rm T}}=10.197\pm0.027$ and 
$V_{{\rm T}}=9.898\pm0.028$, which transforms into $B=10.13$
and $V=9.87$, again compatible with no changes at all.

A more obvious complication for this model is that the metallic lines
do not seem to be any narrower than other presumably photospheric lines.
As a matter of fact, it is difficult to estimate the rotational velocity
of the star. Using the linear relationships for the width of \ion{He}{i}
lines developed by Steele et al. (1999) for Be stars observed with the
same configuration, we obtain 
$v\sin i \approx 300\pm40\:{\rm km}\,{\rm s}^{-1}$ from the lines on the
August 1998 spectrum (i.e., that with the weakest metallic spectrum). 
This measurement clearly confirms that the star is a fast rotator, but
there is a large
dispersion in the values estimated from different lines. 

Another problem faced by this interpretation comes from the asymmetry
of the apparent shell lines. It is well known that in Be stars, symmetric
lines arise from a quasi-Keplerian decretion disc, while asymmetric profiles
correspond to perturbed configurations of such a disc (Hanuschik et al. 1988;
1995;1996). Whenever asymmetric lines are seen, the ratio between the
Violet and Red peaks (V/R ratio) varies quasi-cyclically, due to the
development of global one-armed density oscillations \cite{oka00}. In 
contrast, the V/R ratio in the H$\alpha$ line of
\object{BD\,$+53^{\circ}$2790} has been $<1$ during 15 years, in spite
of the large changes in the overall shape and strength of the line. At
the same time, the symmetry of H$\beta$ and \ion{H}{i}
$\lambda$6678\AA\ has changed. This behaviour has not been observed in
other Be stars and makes it unlikely that
\object{BD\,$+53^{\circ}$2790} might be a classical Be star. 
Further arguments against the interpretation of
\object{BD\,$+53^{\circ}$2790} as a classical Be star will be
discussed in a forthcoming paper. 

One further complication is the detection of a likely $\approx 9.5\:{\rm d}$
period. It is difficult to see how an extended envelope could form with such
a close binary companion. The only possibility would be that, due to 
tidal truncation by the companion \cite{NO01}, a very
dense and small shell would form around the Oe star.

\subsection{\object{BD\,$+53^{\circ}$2790} as a spectroscopic binary}

One obvious possibility to explain the presence of a 
lower-temperature spectral component is that the star is in reality an 
unresolved spectroscopic binary, containing an O9 star and a slightly later
companion. If this was the case, the companion should be of spectral type
close to B1 in order to display the \ion{O}{ii} spectrum.
Since the UV spectrum is clearly dominated by a main-sequence O-star,
this companion cannot have a high luminosity and should be fainter than
the O star in the UV and have a comparable magnitude in the $B$ band.

Such a combination could be obtained with, for example, an 
O9V((f))+B1III binary. The O9V star, due to its Of nature, would contribute
the emission in \ion{He}{ii} and an asymmetric component to H$\alpha$. The 
B1 giant could be a Be star and contribute the Balmer and \ion{He}{i}
line emission. It is clear that such configuration could never
produce the observed X-ray luminosity because of colliding winds, since
both stars would have relatively weak winds and even binaries with
very strong winds produce lower luminosities. Moreover, the X-ray spectrum
observed is very different from those of colliding wind systems, which
are generally soft (with little signal above 10 KeV) and interpreted as 
coming from hot,
optically thin plasma with fixed or variable solar element abundances
(Skinner et al. 1998; Stevens et al. 1996). None of such models or a
combination of them (MEKAL, VMEKAL, RAYMOND in {\sc xspec} terminology) 
gave acceptable fits to the X-ray spectrum of \object{4U\,2206+54}. 
Therefore, the presence of a compact object is necessary. Given the 
$\approx 9.5\:{\rm d}$ period, the compact object would be orbiting 
one of the components (presumably the Of star, since the Be star would
not have a wind that could explain the observed X-ray lightcurve) in 
a close orbit and accreting from its wind. The other component 
would then be in a much wider orbit. 

 Abt \& Bautz (1963) found no evidence for binarity in 
\object{BD\,$+53^{\circ}$2790} in their study of radial velocities of
early-type stars. The reported value $v_{{\rm rad}} = -49.5\:{\rm
  km}\,{\rm s}^{-1}$ is typical for a 
member of the Perseus arm. However, this does not rule out the binary
model, since the system could be very wide. Accurate radial velocity
measurements of the weak metallic lines in order to check whether they
are consistent with the strongest lines, should be a test of this
hypothesis.

\subsection{\object{BD\,$+53^{\circ}$2790} as a single peculiar active star.}

The spectral variability of \object{BD\,$+53^{\circ}$2790} resembles
in many aspects that of He-rich stars, though these objects have later
spectral types (clustered tightly around B2). Some of these systems
are known to display spectral
variability in their \ion{Si}{iii} lines in antiphase with their \ion{He}{i}
lines. These variations are strictly cyclical and correspond to
the rotational period of the star, which is $\sim 1\:{\rm d}$ 
\cite{wal82}. Some of the He-rich
stars (allegedly, those which are fast rotators) display H$\alpha$ in 
emission, but the line profile does not look shell-like \cite{zboal97}.
Only the peculiar He-rich star \object{$\sigma$ Ori E} displays an
H$\alpha$ line profile similar to that of
\object{BD\,$+53^{\circ}$2790}, but in this star the V/R ratio varies
cyclically with the same period as the \ion{He}{i} lines. In any case,
\object{BD\,$+53^{\circ}$2790} is unlikely to be related to He-rich
stars (say, as an early-spectral type relative), because all the
values of EWs measured for \ion{He}{i} lines in
\object{BD\,$+53^{\circ}$2790} are compatible with absolutely no
changes in their strength or the \ion{He}{i}/\ion{H}{i} ratio.

In spite of this, the possibility that the peculiar changes seen in the
spectrum of \object{BD\,$+53^{\circ}$2790} can be attributed to some
unknown physical mechanism operating in a single star remains.

\section{Discussion}

Whatever the model adopted for \object{BD\,$+53^{\circ}$2790}, it is
clear that its ultraviolet and blue spectrum is dominated by a star of
approximate spectral type O9.5V. For such a star, the intrinsic colour
would be close to $(B-V)_{0} = -0.3$. Then, using the value from St84,
$(B-V)=0.2$, we have $E(B-V)\approx0.5$ and, assuming standard reddening, 
$A_{V}=R\times E(B-V)=1.6$.
 An O9.5V star has an absolute magnitude $M_{V} = -4.3$ (Vacca et al. 1996)
and therefore the measured $V=9.85$ implies $d\approx3\:{\rm kpc}$.
At this distance, the average X-ray luminosity of \object{4U\,2206+54} is 
$L_{{\rm x}} \ga 10^{35}\:{\rm erg}\,{\rm s}^{-1}$.

The observational history of \object{4U\,2206+54}, which 
has been detected by all satellites that
have pointed at it and has never been observed to undergo an
outburst, is notably different from that of Be/X-ray transients,
such as 4U\,0115+63 or A\,0535+26 (see Negueruela \& Okazaki 2000). 
There is a second
subclass of Be/X-ray binaries characterised by low-luminosity
persistent X-ray emission with little variation, e.g., \object{X
Persei} \cite{hab98}, but they are believed to have large orbital
periods \cite{RR99} -- which is certainly the case for \object{X
Persei} \cite{del01}. The 9.5-d period of \object{4U\,2206+54} makes a
connection unlikely, though the low X-ray luminosity Be/X-ray binary
\object{3A\,0726$-$26} ($P_{\rm s}=103.2\:{\rm s}$) could have a 
short period  $P_{\rm orb}=34.5\:{\rm d}$ \cite{CP97}. Moreover the
decrease in $L_{{\rm x}}$ by more than one order of magnitude
reported by SA92 is also atypical of these systems.
    
The X-ray luminosity and behaviour, with short erratic flares, point then
to wind accretion as the mechanism producing the X-rays, if the compact 
object is a neutron star. The strong wind
profiles seen in the UV spectrum indicate a large mass-loss rate which 
would fuel the X-ray system. Because of this, we can expect a
  similarity with low-luminosity SXBs, at least as long as the
  accretion process is concerned.  Wind accreting supergiants with
  orbital periods similar to \object{4U\,2206+54} generally show
  rather higher luminosities. This is certainly the case of
  \object{Vela X-1} ($P_{\rm orb}=8.96\:{\rm d}$) with 
$L_{{\rm x}} \approx 4\times10^{36}\:{\rm erg}\,{\rm s}^{-1}$
\cite{kre99} and likely \object{2S\,0114+65} ($P_{\rm orb}=11.6 \: {\rm
  d}$) if the high distance estimates are correct (e.g., Reig et
al. 1996). Such difference would be due to the
much weaker wind of the Of star when compared to a supergiant.

In this respect, it must be noted that the absorption column to
\object{4U\,2206+54} derived from our X-ray spectral fitting  
$N_{\rm H}= 4.7\pm0.2 \times 10^{22}$ atoms cm$^{-2}$ (not very different
from the values found by SA92) is much larger than what corresponds to the
interstellar absorption. The standard relation from Bohlin et al. (1978)
indicates that $E(B-V)\approx0.5$ translates into 
$N_{\rm H} \approx 3\times 10^{21}$ atoms cm$^{-2}$, i.e., one order
of magnitude less than observed. This would be indicating the presence
of very optically thick material in the vicinity of the 
compact object, though it is not clear how this interpretation
  can be reconciled with
the absence of an iron line.

It is also worth mentioning that the X-ray source \object{RX
  J1826.2$-$1450}, whose optical 
counterpart is also a main-sequence O star, could harbour
a black hole, since it contains a microquasar \cite{paral00},
and has an X-ray luminosity similar to or lower than
  \object{4U\,2206+54}. Therefore
we cannot rule out the possibility of a black hole companion in
\object{4U\,2206+54}, though the presence of a 
high-energy cutoff in the X-ray spectrum (which is typical of X-ray 
pulsars) favours a neutron star companion.

Even though we have no explanation for the spectral changes shown by 
\object{BD\,$+53^{\circ}$2790}, a very close similarity to
\object{2S\,0114+65} is suggested.
The optical counterpart to this system, {\object{V662~Cas}, apparently
  is a normal B1 supergiant \cite{reial96}, in spite of the fact
that some authors have claimed that the strength of the Balmer
lines is not as large as expected for such a  star. However, van Kerkwijk
\& Waters (1989) report an instance of spectral change in this source that 
occurred on November 4th, 1986. On this occasion, the complete metallic
spectrum (i.e., \ion{O}{ii} + \ion{Si}{iii}) of the source -- which 
is typical of a B1 supergiant -- disappeared,
leaving behind what looked like a normal B2-3III spectrum. 
It is interesting that the same set of lines that are variable in
\object{BD\,$+53^{\circ}$2790} also varied in \object{V662 Cas},
though in the latter case, their 
disappearance seems to leave behind a cooler stellar spectrum. We take
this as a suggestion that some stars in binaries with close compact
object companions may be structurally unstable, perhaps due to
their previous history, though at the moment we are unable to
propose any physical mechanism for this variability.

Moreover Guarnieri et al. (1991) and Minarini et al. (1994) report the 
occurrence of optical outbursts in both \object{BD\,$+53^{\circ}$2790}
and \object{V662 Cas},
during which some lines which are generally not seen or in absorption
go strongly into emission and argue that these events suggest that both
stars are Be stars. The classification spectra of both objects show that
they are not classical Be stars, after all, but such episodes point to
a further connection between the two systems. In this respect, it must
be noted that Hall et al. (2000) have presented evidence strongly
suggesting that the 2.7-h periodicity observed in the X-ray lightcurve
of \object{2S\,0114+65} 
should correspond to the spin period of a very slowly rotating neutron star.
If \object{4U\,2206+54} exhibits periodic behaviour on a similar
timescale, it is very unlikely that our observations could have
detected it. 

Whichever of the scenarios proposed turns out to be closer to reality,
the X-ray emission from \object{4U\,2206+54} seems certain to be due to 
direct accretion from the wind of an active O-type star. This adds
to increasing evidence that the traditional divisions of Supergiant
X-ray binaries and Be/X-ray binaries are not enough to describe the
whole set of Massive X-ray binaries. While the case of some objects, like
\object{2S\,0114+65} and \object{4U\,1907+09}, which seem to share
characteristics of both groups, had always shown to be problematic, it
is clear now that there is a variety of objects, such as
\object{RX$\:$J1826.2$-$1450},  
\object{RX$\:$J0050.7$-$7316} and \object{4U\,2206+54}, which simply
do not belong to any of those categories.

\section{Conclusion}

We have presented X-ray observations of the High Mass X-ray binary 
\object{4U\,2206+54}. The erratic flaring, lack of pulsations and
  very high hydrogen column density strongly suggest that the X-ray
emission is produced by accretion from a wind. Though
circumstantial evidence supports the idea that the compact companion
is a neutron star, the lack of pulsations and  an X-ray luminosity 
comparable to that of \object{RX J1826.2$-$1450}, whose optical counterpart is
a main-sequence star and contains a microquasar, leave the possibility
of a black hole companion open. Optical and ultraviolet spectroscopy
of the optical component \object{BD$\:+53^{\circ}\,$2790} show it to
be a very peculiar object, displaying emission in \ion{H}{i},
\ion{He}{i} and \ion{He}{ii} lines and variability in the intensity of
many metallic lines. Strong wind troughs in the UV resonance
  lines suggest a large mass loss rate. These
properties might indicate that the star displays at the same time the
Of and Oe phenomena or even hint at the possibility
that it could be a spectroscopic binary consisting of two massive
stars in addition to the compact object.
There is with all certainty an O9.5V star in the system, which is probably
a mild Of star, and which likely feeds the compact object with its stellar
wind.

\begin{acknowledgements} 

The INT is operated on the island of La Palma by the Royal
Greenwich Observatory in the Spanish Observatorio del Roque de
Los Muchachos of the Instituto de Astrof\'{\i}sica de Canarias. The
G.~D.~Cassini telescope is operated at the Loiano Observatory by the 
Osservatorio Astronomico di Bologna. 
This research has made use of data 
obtained through the High Energy Astrophysics Science Archive Research 
Center Online Service, provided by the NASA/Goddard Space Flight Center
and of the Simbad data base, operated at CDS,
Strasbourg, France.

We would like to thank the referee, Frank Haberl, for his helpful
comments. IN would like to thank Prof. Nolan 
Walborn for useful discussions on the spectral classification and
Dr. Manfred Pakull for valuable comments on the manuscript. The August
1998 spectrum was taken by Dr. I.~A.~Steele. During part
of this work IN was supported by a PPARC fellowship and later by an
ESA external fellowship. PR acknowledges support from the European Union 
through the Training
and Mobility of Researchers Network Grant ERBFMRX/CT98/0195.

\end{acknowledgements}

\end{document}